\documentclass[prb,groupedaddress,superscriptaddress]{revtex4}
\usepackage{graphicx}
\usepackage{amssymb}
\usepackage{latexsym}
\usepackage[usenames]{color}
\usepackage{float}

\begin{document}

\title{Tuning the spin dynamics of kagome systems}

\author{D Wulferding}
\affiliation{Institute for Condensed Matter Physics, Technical University of Braunschweig, D-38106 Braunschweig, Germany}
\author{P Lemmens}
\affiliation{Institute for Condensed Matter Physics, Technical University of Braunschweig, D-38106 Braunschweig, Germany}
\author{H Yoshida}
\affiliation{National Institute for Materials Science, 1-1 Namiki, Tsukuba, Ibaraki 305-0044, Japan}
\author{Y Okamoto}
\affiliation{Institute for Solid State Physics, University of Tokyo, Kashiwa, Chiba 277-8581, Japan}
\author{Z Hiroi}
\affiliation{Institute for Solid State Physics, University of Tokyo, Kashiwa, Chiba 277-8581, Japan}

\begin{abstract}
Despite the conceptional importance of realizing spin liquids in solid states only few compounds are known. On the other side the effect of lattice distortions and anisotropies on the magnetic exchange topology and the fluctuation spectrum are an interesting problem. We compare the excitation spectra of the two $s=1/2$ kagome lattice compounds volborthite and vesignieite using Raman scattering. We demonstrate that even small modifications of the crystal structure may have a huge effect on the phonon spectrum and low temperature properties.

\end{abstract}

\pacs{75.10.Jm, 75.10.Kt, 75.50.Ee}
\maketitle

\section{Introduction}
The suggestion of a resonating valence bond ground state in spin liquid systems together with possible exotic, fractional spinon excitations has triggered an intense research interest.~\cite{anderson, lhuillier-review, balents-review} One of the most promising systems for studying these phenomena is the $s=1/2$ Heisenberg antiferromagnet on the kagome lattice. The ground states proposed by theory for this system include a valence bond crystal~\cite{singh} or a resonating valence bond with a gapped~\cite{yan} or gapless~\cite{ran} excitation spectrum.
In several theoretical studies the Raman response of spin liquids on the kagome lattice has been investigated. The spectrum can consist of low energy modes as well as high energy, broad continua, depending on details of the systems.~\cite{laeuchli, ko} The anisotropy of these signals with respect to the orientation of the electric fields of incident and scattered light can be used to probe local symmetry breaking of the spin degrees of freedom.~\cite{raman-cepas}

Up to now, experimental realizations of the kagome lattice antiferromagnet are plagued by impurities, interlayer coupling, and/or anisotropic magnetic coupling strengths. It is therefore of importance to understand the influence of these shortcomings on the system's ground state.

In both title compounds volborthite (Cu$_3$V$_2$O$_7$(OH)$_2$$\cdot$2H$_2$O) and vesignieite (BaCu$_3$V$_2$O$_8$(OH)$_2$) the kagome planes are slightly distorted, resulting in two inequivalent Cu sites. Such a reduced symmetry can either enhance or result from a spin gap, from dimerization, or from some kind of long range order.

For volborthite, this distortion leads to two different magnetic exchange interactions $J_1$ and $J_2$, with an anisotropy of about 20$\%$ and an average value of $J_{volb} = 86$ K, as determined from magnetic susceptibility and specific heat measurements.~\cite{yoshida09} In magnetic susceptibility measurements a broad maximum is observed around 22 K that points to the onset of short-range magnetic order. Neutron scattering experiments detect nearest neighbor correlations settling for temperatures below $T = 50$ K.~\cite{nilsen}
Finally, $^{51}$V-NMR reveals a magnetic transition around $T = 0.9$ K, with anomalously slow energy spin fluctuations persisting down to lowest temperatures.~\cite{yoshida-prl}

The kagome planes in vesignieite are less distorted, with a difference in bond length of only 0.07$\%$.~\cite{colman} The magnetic exchange interaction is determined as $J_{vesi} = 53$ K. Bulk susceptibility measurements reveal a broad maximum around 21 K.~\cite{okamoto}
The estimated Dzyaloshinsky-Moriya (DM) interaction is around 0.14 $J$, i.e. larger than the quantum critical value of 0.1 $J$~\cite{cepas} and thus the ground state of vesignieite should be in the N\'{e}el-ordered regime.~\cite{colman} Indeed, $^{51}$V NMR shows a partial spin freezing at $T_c = 9$ K.~\cite{quilliam}

So far, herbertsmithite (ZnCu$_3$(OH)$_6$Cl$_2$) is considered to be the closest experimental realization of a structurally perfect $s=1/2$ Heisenberg antiferromagnet on a kagome lattice.~\cite{shores} However, this compound shows antisite disorder with up to 10$\%$ of the magnetic Cu$^{2+}$ ions exchanged by Zn$^{2+}$ ions.~\cite{shlee} Thereby, a weak interlayer magnetic coupling as well as magnetic vacancies are induced in the kagome layers. This might alter the system's magnetic properties from those of a perfect kagome lattice. This possible drawback is avoided in both volborthite and vesignieite, as their intermediate layer is comprised of V$^{5+}$ ions. Thus, antisite disorder of the Cu ions is prevented.

In this brief report, we present inelastic light scattering (Raman scattering) experiments on powder samples of volborthite and vesignieite. Raman scattering is a sensitive probe of lattice and electronic/magnetic degrees of freedom. In particular, Raman scattering has been highlighted as a powerful tool to distinguish possible ground states with weak order or local order parameters.~\cite{raman-cepas, ko} Recently, it has been successfully applied to characterize the excitation spectrum of the spin liquid state in herbertsmithite.~\cite{wulferding}

\section{Experimental}
Powder samples of the title compounds were prepared as described previously.~\cite{hiroi01}
Raman scattering experiments have been performed in quasi backscattering geometry using a frequency-doubled $\lambda = 532$ nm Nd:YAG laser with a fixed laser power of 0.5 mW. Such a small power level is essential to avoid an overheating of the powder samples with reduced thermal conductivity. The samples were installed into a He-cooled closed cycle cryostat with a temperature range of 3 -- 300 K. The spectra were collected via a Dilor-XY 500 triple spectrometer by a liquid nitrogen cooled HORIBA Jobin Yvon CCD (Spectrum One CCD-3000 V).

\section{Results and discussion}
Both compounds volborthite and vesignieite belong to the monoclinic space group $C2/m$. A symmetry analysis has been carried out. It yields the following Raman active modes for volborthite: $\Gamma_{Raman} = 16 A_g + 11 B_g$ and for vesignieite: $\Gamma_{Raman} = 13 A_g + 8 B_g$. 

All presented measurements are performed in ($xx$) polarization, i.e. with parallel incoming and scattered light. The crystallites in the powder sample are semi-transparent and partially scramble the light polarization upon scattering. Polarization dependent experiments not shown here did not lead to additional information. In fact, the spectra obtained in crossed ($yx$) polarization are identical to the ($xx$) spectra. Both compounds show modes around 800 cm$^{-1}$ and 3400 cm$^{-1}$, due to O-H vibrations. The energy of these modes is much larger than the magnetic exchange interaction. Therefore interactions do not exist and they will not be further discussed here.

\begin{figure}
\label{figure1}
\centering
\includegraphics[height=250px]{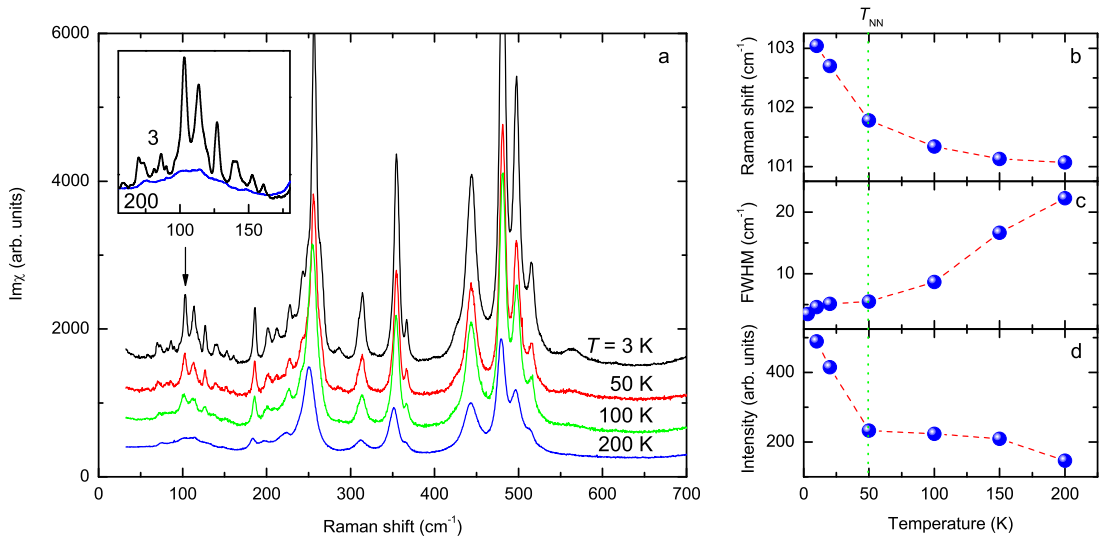}
\caption{(Colour online) a) Bose-corrected Raman spectra of volborthite at $T =$3, 50, 100 and 200 K in ($xx$) polarization; the spectra are shifted in intensity for clarity. b) -- d) Plots of frequency, linewidth and intensity of the phonon mode at 103 cm$^{-1}$ (see arrow) as function of temperature. The inset zooms into the low energy region for spectra obtained at 200 and 3 K.}
\end{figure}

In Figure 1 (a) Raman spectra of volborthite are shown at different temperatures, ranging from 200 to 3 K. With decreasing temperature, the intensity of the phonons increases. Furthermore, in the low energy range below $\approx 175$ cm$^{-1}$, a set of new, sharp phonons appears (see inset). The temperature evolution of one phonon at 103 cm$^{-1}$ is showcased in Figure 1. The energy scale of this phonon group is in the range of 1$J_{volb}$ -- 3$J_{volb}$.

We observe a dramatic gain of intensity for temperatures below 50 K. Around this temperature, the onset of nearest neighbor spin-spin correlations is stated from magnetic susceptibility. At low temperatures, the total number of phonons exceeds the expected 27 modes. It is suggested that via spin-phonon coupling a structural distortion is gradually taking place in volborthite with decreasing temperature. Through this moderate distortion, frustration is releaved. This results in an increase in Raman active phonons that are energetically in the vicinity of $J_{volb}$.

\begin{figure}
\label{figure2}
\centering
\includegraphics[height=250px]{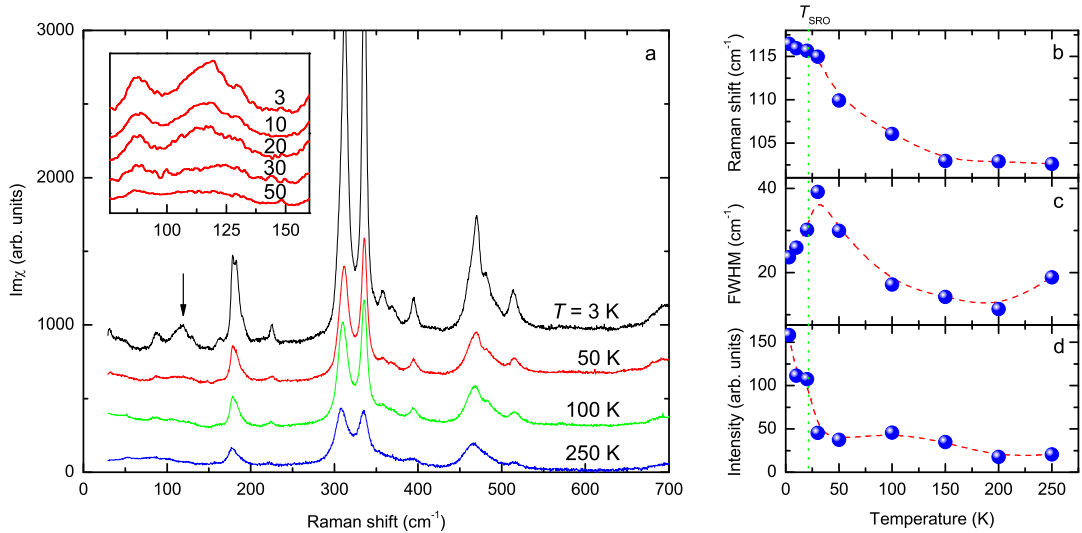}
\caption{(Colour online) a) Bose-corrected Raman spectra of vesignieite at $T =$ 3, 50, 100 and 250 K in ($xx$) polarization; the spectra are shifted in intensity for clarity. (b) -- (d) Plots of the temperature development of the broad mode at 115 cm$^{-1}$ (see arrow). The inset shows the spectral region of magnetic scattering for the temperature range 50 -- 3 K.}
\end{figure}

Figure 2 (a) plots the Raman spectra of vesignieite at different temperatures, in analogy to Figure 1 (a). The phonon modes sharpen and gain in intensity with decreasing temperature, comparable to the spectral changes in volborthite. In the low energy range a broader mode appears, extending from about 95 cm$^{-1}$ to 140 cm$^{-1}$, or roughly 2.6$J_{vesi}$ -- 3.8$J_{vesi}$ (see inset). Its parameters are plotted in Figure 2 (b) -- (d). This mode has a different, non-lorentzian line shape compared to the phonons. Following the temperature development of the mode's parameters, it becomes pronounced below about 25 K, which is close to $T_{SRO} = 22$ K, where short-range magnetic correlations set in.
In the structurally perfect kagome lattice compound herbertsmithite, a broad scattering continuum is observed at low temperatures, ranging from lowest energies up to $6 \cdot J$.~\cite{wulferding} This continuum is attributed to spinon excitations from the resonating valence bond ground state.
The mode in vesignieite is less broad, with an onset at a finite energy. Its maximum position around 115 cm$^{-1}$ corresponds to $3.1 \cdot J_{vesi}$, with $J_{vesi} = 53$ K.~\cite{okamoto} This is in very good agreement with the estimation of a typical energy for magnetic Raman scattering in a Heisenberg antiferromagnet~\cite{cottam} at $E_{2M} = 2J(zS-1) = 110$ cm$^{-1}$, where $z=5$ is the number of nearest neighbors and $s=1/2$ is the spin. Therefore, we attribute this signal to a magnetic scattering process.
The analogous estimation for volborthite leads to $E_{2M} = 180$ cm$^{-1}$. In this energy range no corresponding signal is observed. It may be masked by phonons.

While neither of the two systems display long range order at finite temperatures, the observation of a spinon continuum comparable to herbertsmithite is absent in both compounds.
Therefore, we can conclude that even a slight anisotropy between $J_1$ and $J_2$ of 0.07$\%$ as in vesignieite has a stronger influence on the spin liquid ground state than antisite disorder in herbertsmithite of about 10$\%$.

\section{Conclusion}

In summary, we investigated the lattice and magnetic excitations in kagome systems. Our results reveal spin-phonon coupling in the moderately distorted volborthite that lowers the crystal symmetry and enhances spin-spin correlations for temperatures below 50 K. Vesignieite, with a close to perfect kagome lattice structure, shows no additional low temperature phonon modes but instead magnetic scattering. Our study demonstrates that the spin dynamics of spin liquid systems depend decisively on the underlying structure.

We would like to thank V. P. Gnezdilov for valuable discussions. This work was supported by DFG, B-IGSM and the NTH School ``Contacts in Nanosystems''.

\end{document}